\documentclass[prl,twocolumn,superscriptaddress,showpacs,amsmath,amssymb]{revtex4-1}
\usepackage[dvipdfmx]{graphicx}
\usepackage{natbib}
\usepackage{bm}
\usepackage{color}
\usepackage{comment}
\usepackage{braket}

\newcommand{\ketbra}[2]{|#1\rangle\langle#2|}
\newcommand{\expect}[1]{\langle #1 \rangle}
\def\H{{\rm H}}
\def\V{{\rm V}}

\def\D{{\rm D}}
\def\A{{\rm A}}
\def\B{{\rm B}}

\def\U#1{{\rm #1}}
\def\u#1{_{\rm #1}}

\begin{document}
\title{Robust entanglement distribution via telecom fibre assisted by an asynchronous counter-propagating laser light}

\author{Koichiro Miyanishi}
\email{miyanishi@qi.mp.es.osaka-u.ac.jp}
\affiliation{Graduate School of Engineering Science, Osaka University,
Toyonaka, Osaka 560-8531, Japan}
\author{Yoshiaki Tsujimoto}
\affiliation{Graduate School of Engineering Science, Osaka University,
Toyonaka, Osaka 560-8531, Japan}
\affiliation{Advanced ICT Research Institute, National Institute of Information and Communications Technology (NICT), Koganei, Tokyo 184-8795, Japan}
\author{Rikizo Ikuta}
\affiliation{Quantum Information and Quantum Biology Division, Institute for Open and Transdisciplinary Research Initiatives, Osaka University, Osaka 560-8531, Japan}
\affiliation{Graduate School of Engineering Science, Osaka University, Toyonaka, Osaka 560-8531, Japan}
\author{Shigehito Miki}
\affiliation{Advanced ICT Research Institute, National Institute of Information
and Communications Technology (NICT), Kobe, Hyogo 651-2492, Japan}
\affiliation{Graduate School of Engineering Faculty of Engineering, Kobe University,
1-1 Rokko-dai cho, Nada-ku, Kobe 657-0013, Japan}
\author{Masahiro Yabuno}
\affiliation{Advanced ICT Research Institute, National Institute of Information
and Communications Technology (NICT), Kobe, Hyogo 651-2492, Japan}
\author{Taro Yamashita}
\affiliation{Advanced ICT Research Institute, National Institute of Information
and Communications Technology (NICT), Kobe, Hyogo 651-2492, Japan}
\affiliation{Japan Science and Technology Agency, PRESTO, Kawaguchi, Saitama 332-0012, Japan}
\author{Hirotaka Terai}
\affiliation{Advanced ICT Research Institute, National Institute of Information and Communications Technology (NICT), Kobe, Hyogo 651-2492, Japan}
\author{Takashi Yamamoto}
\affiliation{Graduate School of Engineering Science, Osaka University, Toyonaka, Osaka 560-8531, Japan}
\affiliation{Quantum Information and Quantum Biology Division, Institute for Open and Transdisciplinary Research Initiatives, Osaka University, Osaka 560-8531, Japan}
\author{Masato Koashi}
\affiliation{Photon Science Center,
The University of Tokyo, Bunkyo-ku, Tokyo 113-8656, Japan}
\author{Nobuyuki Imoto}
\affiliation{Quantum Information and Quantum Biology Division, Institute for Open and Transdisciplinary Research Initiatives, Osaka University, Osaka 560-8531, Japan}

\begin{abstract}
Distributing entangled photon pairs over noisy channels is an important task for various quantum information protocols.
Encoding an entangled state in a decoherence-free subspace (DFS) formed by multiple photons is a promising way to circumvent the phase fluctuations and polarization rotations in optical fibres. 
Recently, it has been shown that the use of a counter-propagating coherent light as an ancillary photon enables us to faithfully distribute entangled photon with success probability proportional to the transmittance of the optical fibres.
Several proof-of-principle experiments have been demonstrated, in which entangled photon pairs from a sender side and the ancillary photon from a receiver side originate from the same laser source.
In addition, bulk optics have been used to mimic the noises in optical fibres.
Here, we demonstrate a DFS-based entanglement distribution over 1km-optical fibre using DFS formed by using fully independent light sources at the telecom band.
In the experiment, we utilize an interference between asynchronous photons from cw-pumped spontaneous parametric down conversion~(SPDC) and mode-locked coherent light pulse.
After performing spectral and temporal filtering, the SPDC photons and light pulse are spectrally indistinguishable.
This property allows us to observe high-visibility interference without
performing active synchronization between fully independent sources.
\end{abstract}

\maketitle

\section{introduction}
Sharing entanglement between two distant parties is an important prerequisite to realize quantum internet~\cite{kimble2008,Wehner18} including quantum key distribution~\cite{Gisin2002,Lo2014}, quantum repeaters~\cite{Sangouard2011} and quantum computation between distant parties~\cite{Cirac1999,Broadbent2009}.
When the sender prepares entangled photon pairs between signal and idler photons and then sends the signal photon to the receiver through an optical fibre, 
the amount of the entanglement becomes smaller or completely destroyed mainly 
due to phase fluctuations and/or polarization rotations in the optical fibre, unless those fluctuations and rotations are actively stabilized.
One of the promising ways to overcome these disturbances is the use of a decoherence-free subspace~(DFS) formed by 
$n$ photons~\cite{Lidar2003}.
So far, many types of DFS-based photon distributions have been experimentally demonstrated~\cite{Kwiat2000,Walton2003,Boileau2004,Bourennane2004,Boileau22004,Yamamoto2005,Chen2006,Prevedel2007,Yamamoto2007,Yamamoto2008}.
A drawback of such DFS protocols is that the efficiency is proportional to $n$-th power of channel transmittance and thus the communication distance is severely limited.
In Ref.~\cite{IkutaDFS2011}, the efficiency of DFS against phase disturbance has been improved to be proportional to the channel transmittance with the use of a counter-propagating laser light for the ancillary photon from the receiver to the sender.
Recently, the method was applied to general collective noise by introducing two independent transmission channels~\cite{Kumagai2013,Ikuta2017}.
In these improved DFS schemes, quantum interference between the idler photon and the coherent light pulse is used.
In practice, the signal photon and the light pulse are expected to be prepared independently by the sender and by the receiver.
However, in the demonstrations~\cite{IkutaDFS2011,Ikuta2017}, the signal photon and the light pulse were prepared from the same laser source as is the case of the original DFS protocols in which both the signal and the ancillary photons are prepared by the sender.
In addition, these experiments are performed in free space with bulk optics simulating the noises and losses of the optical fibres.

In this paper, we report DFS-protected entanglement distribution through optical fibres by preparing the signal and the ancillary photons from fully independent telecom light sources.
We use entangled photon pairs prepared by cw-pumped spontaneous parametric down-conversion (SPDC) 
and the ancillary light pulse prepared by a mode-locked laser.
In general, precise spectral-temporal mode matching is necessary for achieving high-visibility quantum interference between independent light sources.
In our system, cw-pumped SPDC with time-resolved measurement~\cite{Tsujimoto2018} achieves temporal mode matching with reference laser without active synchronization.
The use of the coherent light pulse realizes spectral mode matching with the cw-pumped SPDC photon by using conventional frequency filters.
This cw-pulse hybrid scheme will be useful for connecting different physical systems at a distance.

\section{Results}
\subsection{Counter-propagation DFS Protocol}
\begin{figure}
    \includegraphics[width=9cm]{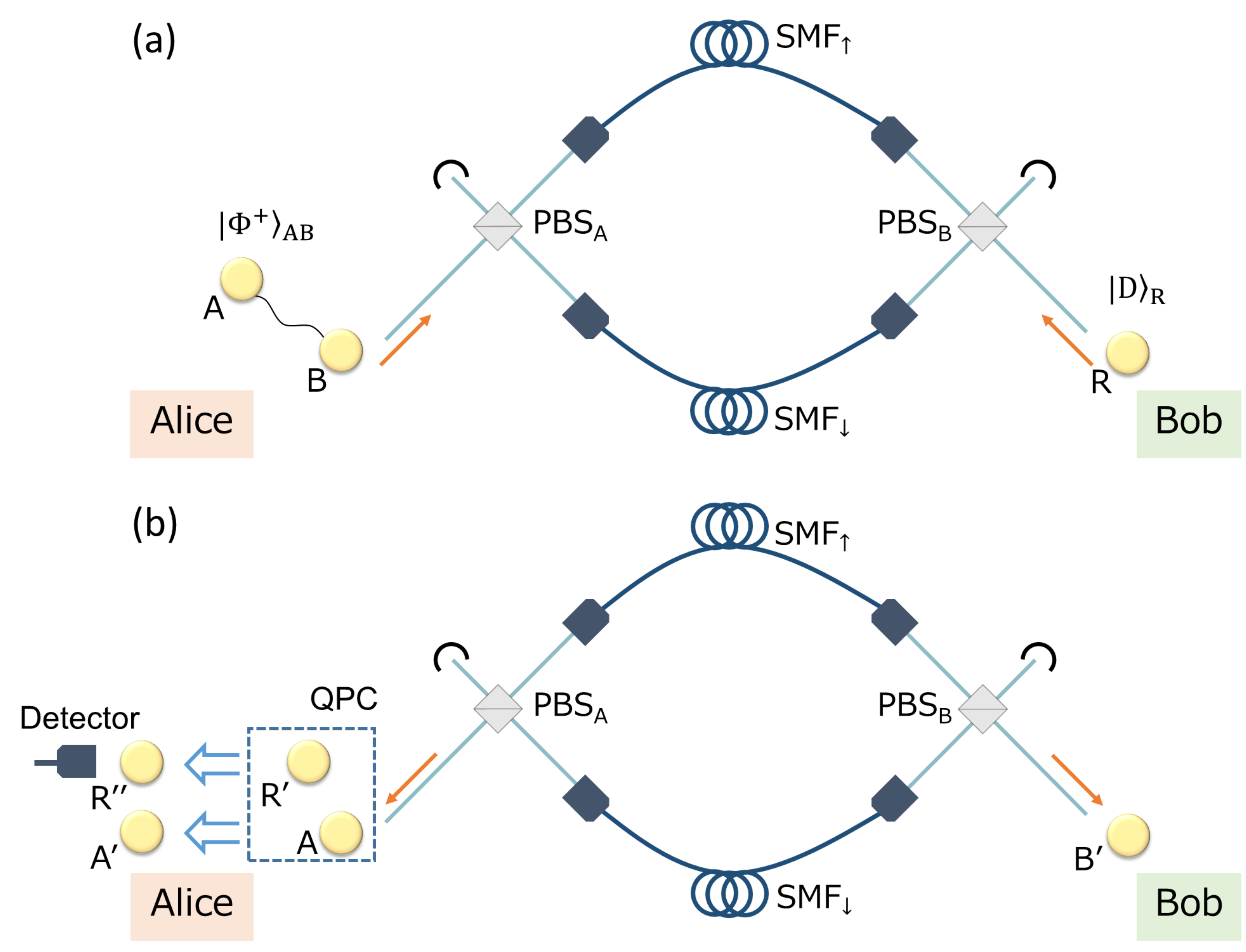}
       \caption{Schematic diagram of entanglement distribution protocol using counter-propagating ancillary photons. (a)Alice prepares $\ket{\Phi^+}_{\rm{AB}}$ and send photon B to Bob. 
       Bob prepares a diagonally-polarized ancillary photon R, and sends it to Alice.  (b)After the transmission of the SMFs, Alice performs the QPC on photons A and R.}
    \label{DFSprotocol}
\end{figure}

In this section, we explain the DFS protocol for sharing an entangled photon pair through two single-mode fibres~(SMFs).
We assume that the SMFs are lossy collective noise channels i.e. 
the noise varies slowly compared to the propagation time of the photons forming the DFS
such that the effect of temporal fluctuation of the noise during that period is negligibly small.
No assumptions are needed for the correlation between the fluctuations in the two SMFs.
As shown in Fig.\ref{DFSprotocol}, first, the sender Alice prepares a maximally entangled photon-pair $\ket{\Phi^+}_{\rm{AB}}=(\ket{\H\H}_{\A\B}+\ket{\V\V}_{\A\B})/\sqrt{2}$, while the receiver Bob prepares an ancillary photon $\ket{\D}_{\rm{R}}$.
Here, $\ket{\H}$, $\ket{\V}$ and $\ket{\D}$ represent horizontal~(H), vertical~(V) and diagonal~(D) polarization states of a photon.
Then, Alice sends photon B to Bob through two SMFs after splitting the H- and V-polarized components of the photon B by using a polarization beamsplitter~(PBS$_{\rm{A}}$).
Bob sends his photon R to Alice in the same way.
After the transmission through the SMFs, the separated H- and V-polarized components of the photons B~(R) are recombined by the $\rm{PBS_{B(A)}}$ at Bob's~(Alice's) side.

From the assumption of the collective noises,
the transformation of the polarization state formed by photons A, B and R in the SMFs is expressed as
\begin{align}
\label{noiseinfibre}
&\ket{\Phi^+}_{\rm{AB}}\otimes\ket{\D}_{\rm{R}}\nonumber\\
&\rightarrow\frac{1}{2}(\alpha^f_{\H,\H}\ket{\H\H}_{\rm{AB}}+\alpha^f_{\H,\V}\ket{\H\V}_{\rm{AB}}\nonumber\\
&\,\,\,\,\,\,\,+\beta^f_{\V,\H}\ket{\V\H}_{\rm{AB}}+\beta^f_{\V,\V}\ket{\V\V}_{\rm{AB}})\nonumber\\
&\,\,\,\,\,\,\,\otimes(\alpha^b_{\H,\H}\ket{\H}_{\rm{R}}+\alpha^b_{\H,\V}\ket{\V}_{\rm{R}}\nonumber\\
&\,\,\,\,\,\,\,+\beta^b_{\V,\H}\ket{\H}_{\rm{R}}+\beta^b_{\V,\V}\ket{\V}_{\rm{R}}),
\end{align}
where $\alpha^{f(b)}_{i,j}$ is the probability amplitude with which the $i$-polarized photon is transformed into the $j$-polarized photon 
through the forward(backward) propagation in the SMF$_{\uparrow}$ and $\beta^{f(b)}_{i,j}$ is the same for SMF$_{\downarrow}$.
In the reciprocal media such as SMFs, $\alpha^f_{\H,\H}=\alpha^b_{\H,\H}:=\alpha_\H$ and $\beta^f_{\V,\V}=\beta^b_{\V,\V}:=\beta_\V$ hold~\cite{Kumagai2013}.
After passing through the SMF$_{\uparrow}$ and SMF$_{\downarrow}$, the H and V polarization photons are respectively extracted from the lower ports of PBS$\u{A}$ and PBS$\u{B}$.
Then, Alice and Bob obtain an unnormalized state as
\begin{align}
\label{afterPBS}
\ket{\Psi}_{\rm{AB^{\prime}R^{\prime}}}&=
\frac{1}{2}( \alpha_\H^2\ket{\H\H\H}_{\rm{AB^{\prime}R^{\prime}}}+\beta_\V^2\ket{\V\V\V}_{\rm{AB^{\prime}R^{\prime}}}\nonumber\\
&\,\,\,\,\,\,\,+\alpha_\H \beta_\V\ket{\H\H\V}_{\rm{AB^{\prime}R^{\prime}}}+\alpha_\H \beta_\V\ket{\V\V\H}_{\rm{AB^{\prime}R^{\prime}}}).
\end{align}
The latter two terms in Eq.(\ref{afterPBS}) are extracted by performing the quantum parity check~(QPC) on photons A and R$'$ in Alice's side~\cite{Pittman2001}.
The Kraus operators $K$ and $\bar{K}$ for the successful and failure events of  the QPC are described by $K=\ket{\H\H}_{\rm{A'R''}}\bra{\H\V}_{\rm{AR'}}+\ket{\V\V}_{\rm{A'R''}}\bra{\V\H}_{\rm{AR'}}$ and $\bar{K}=\sqrt{I-K^\dagger K}$, respectively.
When Alice performs projective measurement $\{\ketbra{\D}{\D}, \ketbra{\A}{\A}\}$ on the photon in mode R$''$, the remaining polarization state shared between Alice and Bob becomes $\ket{\Phi^+}_{\rm{A'B^{\prime}}}$ or $\ket{\Phi^-}_{\rm{A'B^{\prime}}}$ according to the measurement result, where $\ket{\Phi^-}_{\rm{A'B^{\prime}}}=(\ket{\H\H}_{\rm{A'B^{\prime}}}-\ket{\V\V}_{\rm{A'B^{\prime}}})/\sqrt{2}$.
$\ket{\Phi^-}$ can be converted to $\ket{\Phi^+}$ by performing phase flip operation on photon A$'$.
The overall successful probability of this scheme is $|\alpha_\H|^2|\beta_\V|^2/2$.

Assuming that the phase shifts and polarization rotations are completely random and the transmittance of a single photon for each mode is $T=|\alpha_\H|^2=|\beta_\V|^2$, the probability with which photon B and R transmit the lossy channels is $\mathcal{O}(T^2)$.
This probability is improved to $\mathcal{O}(T)$ by using coherent light pulse with an average photon number of $\mu T^{-1}$ at Bob's side instead of the single ancillary photon~\cite{Kumagai2013, IkutaDFS2011}.
In the experiment, the entangled photon pairs are prepared by the spontaneous parametric conversion~(SPDC) with the photon pair generation probability $\gamma$.
At this average photon number, the condition for suppressing the accidental coincidence events caused by the multi-photon emission is $1\gg\mu\gg\gamma$~\cite{Ikuta2017, IkutaDFS2011}.

\subsection{Experimental setup}

\begin{figure*}
\includegraphics[width=18cm]{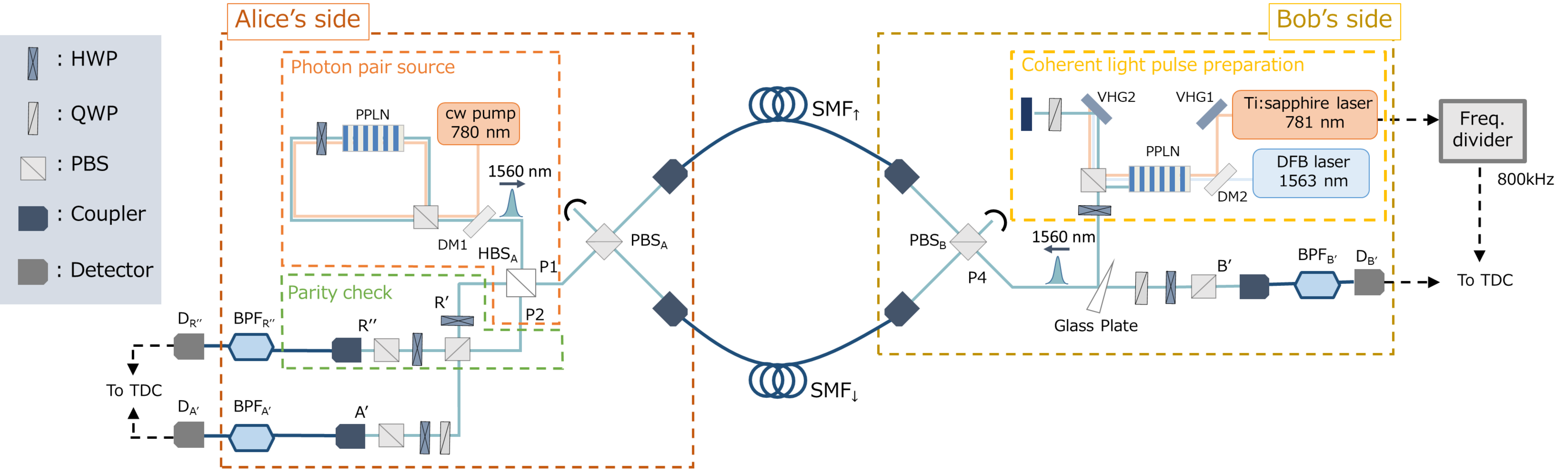}
\caption{The experimental setup for our DFS protocol. At Alice's photon pair source, the cw pump light at 780~nm for SPDC is prepared by second-harmonic generation of the light at 1560~nm from an external cavity diode laser with a linewidth of 1.8~kHz.
At Bob's ancillary photon source, the pulsed light at 781~nm is filtered by a volume holographic grating~(VHG1) with a bandwidth of 0.3~nm and the pulsed light and the cw pump light at 1563~nm are combined by the DM2 and coupled into a PPLN/W.
At the data collection stage by TDC, the repetition rate of the electric signal from the Ti:S pulse laser is divided into 800~kHz to reduce the amount of data for avoiding the data overflow.}
\label{DFSsetup}
\end{figure*}

The experimental setup is shown in FIG.~\ref{DFSsetup}.
At Alice's side, an entangled photon pair in $\ket{\Phi^+}$ at 1560~nm is generated by the SPDC.
For this, a periodically-poled lithium niobate waveguide~(PPLN/W) is placed in the Sagnac interferometer with a PBS~\cite{Tsujimoto2017}, and 7-mW cw light at 780~nm with a diagonal polarization is injected as pump light.
The pump light is removed from the SPDC photons by a dichroic mirror~(DM1).
At a half beam splitter~(HBS$\rm{_A}$), the SPDC photons are divided into two different spatial paths P1 and P2 with probability 1/2
and the entangled photon pair in $\ket{\Phi^+}$ is prepared.
We call the photon in path P2 as photon A and the photon in path P1 as photon B.
We note that there are cases where the SPDC photon pair take the same path P1 or P2, 
but these events are removed by coincidence measurements which are described later.

Photon A is fed to the QPC circuit which we describe later.
H- and V-polarized components of the photon B are divided by $\rm{PBS_A}$ and sent to Bob through 1km $\rm{SMF_{\uparrow}}$ and $\rm{SMF_{\downarrow}}$.
After passing through the SMFs, they are recombined at $\rm{PBS_B}$.
The photon in mode B$'$ extracted from the lower port of $\rm{PBS_B}$ goes to photon detector $\rm{D_{B'}}$.

At Bob's side, a weak coherent light pulse at 1560~nm is prepared for an ancillary photon R by difference frequency generetion~(DFG) at another PPLN/W.
The signal light for DFG comes from a Ti:sapphire~(Ti:S) laser at 781~nm (pulse width of 100~fs, and repetition rate of 80~MHz) after a volume holographic grating~(VHG1) with a bandwidth of 0.3~nm.
The cw pump light at 1563~nm is prepared by DFB laser amplified to a power of 120~mW.
After the DFG process, the 1563~nm pump light and the remaining 781~nm light are separated from the converted 1560~nm light by VHG2 with a bandwidth of 1~nm.
The DFG light is set to a diagonal polarization by a HWP and sent to Alice in the same way as photon B through the two SMFs after reflected by a glass plate~(GP) with a reflectance of 5~\%.
After the transmission, the light pulse from the lower port of the PBS$_{\rm{A}}$ passes through the HBS$\rm{_A}$, and then goes to the QPC circuit.

At the QPC circuit, the H and V component of the coherent light pulse for photon R$'$ are flipped by the HWP, and the light meets photon A at the PBS.
Alice projects the photons in the output mode R$''$ of the PBS onto the diagonal polarization by a HWP and a PBS.
Finally, when Alice postselects the cases where at least one photon is detected at D$\u{R''}$ and D$\u{A'}$ and Bob postselects the cases where at least one photon is detected at D$\u{B'}$,
a maximally entangled state $\ket{\Phi^+}$ is shared between the modes A$'$ and B$'$.
We used superconducting nanowire single photon detectors~(SNSPDs) for D$\u{A'}, $D$\u{B'}$ and D$\u{R''}$.
The timing jitter of these SNSPDs is 85~ps each~\cite{Miki2013}.
To attain a high fidelity QPC operation, precise spectral-temporal mode matching between the heralded cw-pumped SPDC photons and the coherent light pulse is required.
Regarding the temporal mode matching, thanks to the asynchronous nature of cw-pumped SPDC photons, we can extract the event that the coherent light pulse and the heralded single photon exists at the same time by postprocessing.
For the postprocessing, the electric signals from $\rm{D_{R''}}$, $\rm{D_{A'}}$, $\rm{D_{B'}}$ and the clock signal from the Ti:S laser
are connected to a time-to-digital converter~(TDC) which collects all of timestamps with time slot of 1~ps.
Regarding the spectral mode matching, the heralded single photon and the coherent light pulse should be filtered by narrow frequency filters.
We inserted frequency filters whose bandwidths are 0.1~nm for mode B$'$ and 0.03~nm for modes A$'$ and R$''$.
The coherence time of SPDC photons and coherent light pulse are measured to be 169~ps and 176~ps.

\subsection{Experimental results}
\begin{figure}
    \includegraphics[width=9cm]{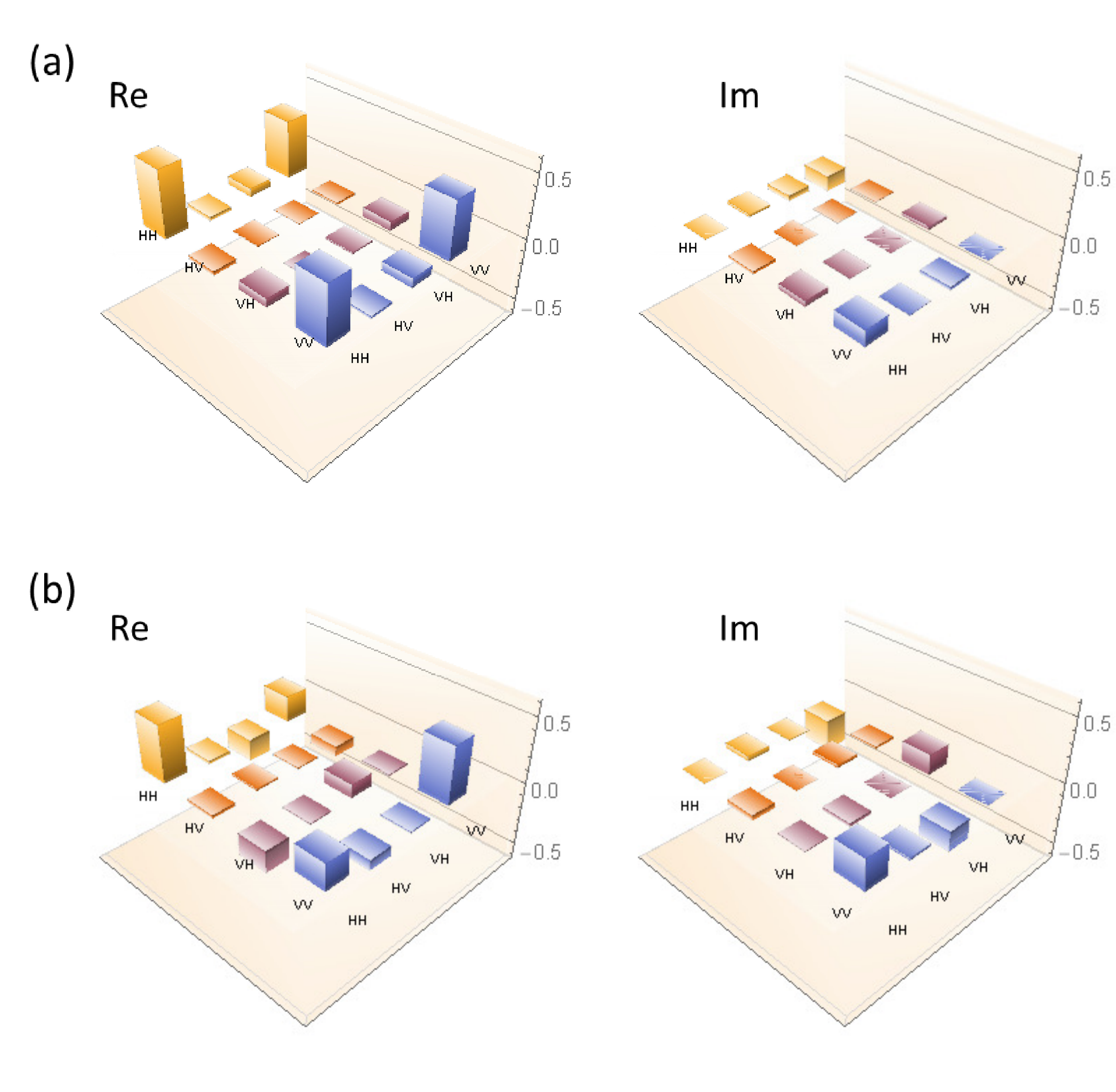}
    \caption{(a)The real part and imaginary part of the $\rho_{\rm{initial}}$. (b)The real part and imaginary part of the $\rho_{\rm{DFS}}$.}
    \label{DM}
\end{figure}

We first characterized the polarization state of the initial photon pair in paths P1 and P2 by performing the quantum state tomography~\cite{James2001} and using the iterative maximum likelihood estimation~\cite{Rehacek2007}.
For this purpose, we inserted a pair of a HWP and a QWP in each output mode just after the HBS$\u{A}$.
The reconstructed density operator $\rho_{\rm{initial}}$ is shown in Fig.~\ref{DM}~(a).
The observed fidelity of $\rho_{\rm{initial}}$ to the maximally entangled state defined by $F =\bra{\Phi^+}\rho_{\rm{initial}}\ket{\Phi^+}$ was $F=0.94\pm0.01$.
This result shows Alice prepares highly entangled photon pairs.

Next, we performed the DFS protocol.
The average photon number $\mu$ of the coherent light pulse after HBS$\rm{_A}$ was set to $\mu \approx 3.1\times10^{-1}$ per window time of 300~ps.
The photon pair generation probability $\gamma$ per coincidence between 300-ps window for idler photons and 100-ps window for signal photons
was measured to be $\gamma \approx 9.0\times10^{-3}$.
The reason for employing large detection widows for $\rm{D_{A'}}$ and $\rm{D_{R''}}$ is to increase the count rate.
$\gamma$ and $\mu$ satisfy the condition $1\gg\mu\gg\gamma$.

Using the three-fold coincidence events among D$\u{R''}$, D$\u{A'}$ and D$\u{B'}$~(see Method), we reconstructed the density operator $\rho_{\rm{DFS}}$ of the photon pair shared between Alice and Bob.
The real part and imaginary part of $\rho_{\rm{DFS}}$ are shown in Fig.~\ref{DM} (b).
The observed fidelity to $\ket{\Phi^+}$ was $F=0.64\pm0.1$.
The maximized fidelity with the local phase shift defined by $F_{\theta} =\rm{max_{-\pi\leq\theta\leq\pi}}\bra{\Phi^+_{\theta}}\rho_{initial}\ket{\Phi^+_{\theta}}$ was $F_{\theta} = 0.75\pm 0.10$ with $\theta=-0.87$ rad, where $\ket{\Phi^+_{\theta}}\equiv\left(\ket{\H\H}+\rm{e}^{i\theta}\ket{\V\V}\right)/\sqrt{2}$.
The result shows that the DFS scheme protects the entanglement against the collective noise in 1~km of SMF.

\section{discussion}

We discuss the reason for the degradation of the fidelity.
Assuming the perfect mode matching between the coherent light pulse and the photon heralded by photon detection at D$\u{B'}$,
we investigate the influence of multiple photons in the coherent light pulse and the multiple photon pair emission of SPDC on the fidelity.
Based on the analysis in Ref.~\cite{Tsujimoto2017},
we construct a simple theoretical model and derive the relation among
the intensity ratio of heralded photon to coherent light pulse,
the intensity correlation function of the heralded photon and the coherent light pulse and 
signal to noise ratio of polarization of the heralded photon.
We introduce the polarization correlation visibilities of the final state defined by 
$V\u{Z}:=|P\u{HH}+P\u{VV}-P\u{HV}-P\u{VH}|/(P\u{HH}+P\u{VV}+P\u{HV}+P\u{VH})$, 
$V\u{X}:=|P\u{DD}+P\u{AA}-P\u{DA}-P\u{AD}|/(P\u{DD}+P\u{AA}+P\u{DA}+P\u{AD})$ 
and $V\u{Y}:=|P\u{RR}+P\u{LL}-P\u{RL}-P\u{LR}|/(P\u{RR}+P\u{LL}+P\u{RL}+P\u{LR})$,
where $P_{mn}$ is the coincidence probability among D$\u{R''}$ with D-polarization, 
D$\u{B}$ with $m$-polarization and D$\u{A'}$ with $n$-polarization.
Here, $m,n$ =H, V, D, A, R, L, where A, R and L
represent antidiagonal polarization, right circular polarization and left circular polarization, respectively.
Then, the fidelity of the final state is given by $F=(1+V\u{Z}+V\u{X}+V\u{Y})/4.$
The explicit formula and detailed derivation are described in the supplemental material.  
We assume that the intensity correlation function of the stray photons is 2 and that of the coherent light pulse is 1.
From the experimental result, the signal to noise ratio of polarization of the heralded single photon is $56$ and the intensity ratio between the heralded single photon and the coherent light pulse is $0.28$.
By running another experiment, the intensity correlation function of the heralded single photon was estimated to be $0.098$. 
Using these parameters, we estimate the theoretical value of fidelity as 0.81.
This value is slightly higher than the experimentally-obtained value to be 0.75.
We guess the gap of 0.06 is caused by the frequency mode mismatch between the signal photon and the coherent light pulse.

The above analysis indicates that the main cause of the degradation of the fidelity stems from multiple pairs in SPDC photons and/or coherent light pulse. 
Decreasing mean photon numbers of the sources leads to the higher fidelity, but the rate of sharing entanglement between Alice and Bob gets smaller.
Introducing frequency filters with higher transmittance, low-jitter photon detectors~\cite{Miki2019}, 
and frequency multiplexing~\cite{Aktas16} will help to decrease pump intensity while keeping the entanglement sharing rate.

\section{Methods}

\begin{figure}
    \includegraphics[width=9cm]{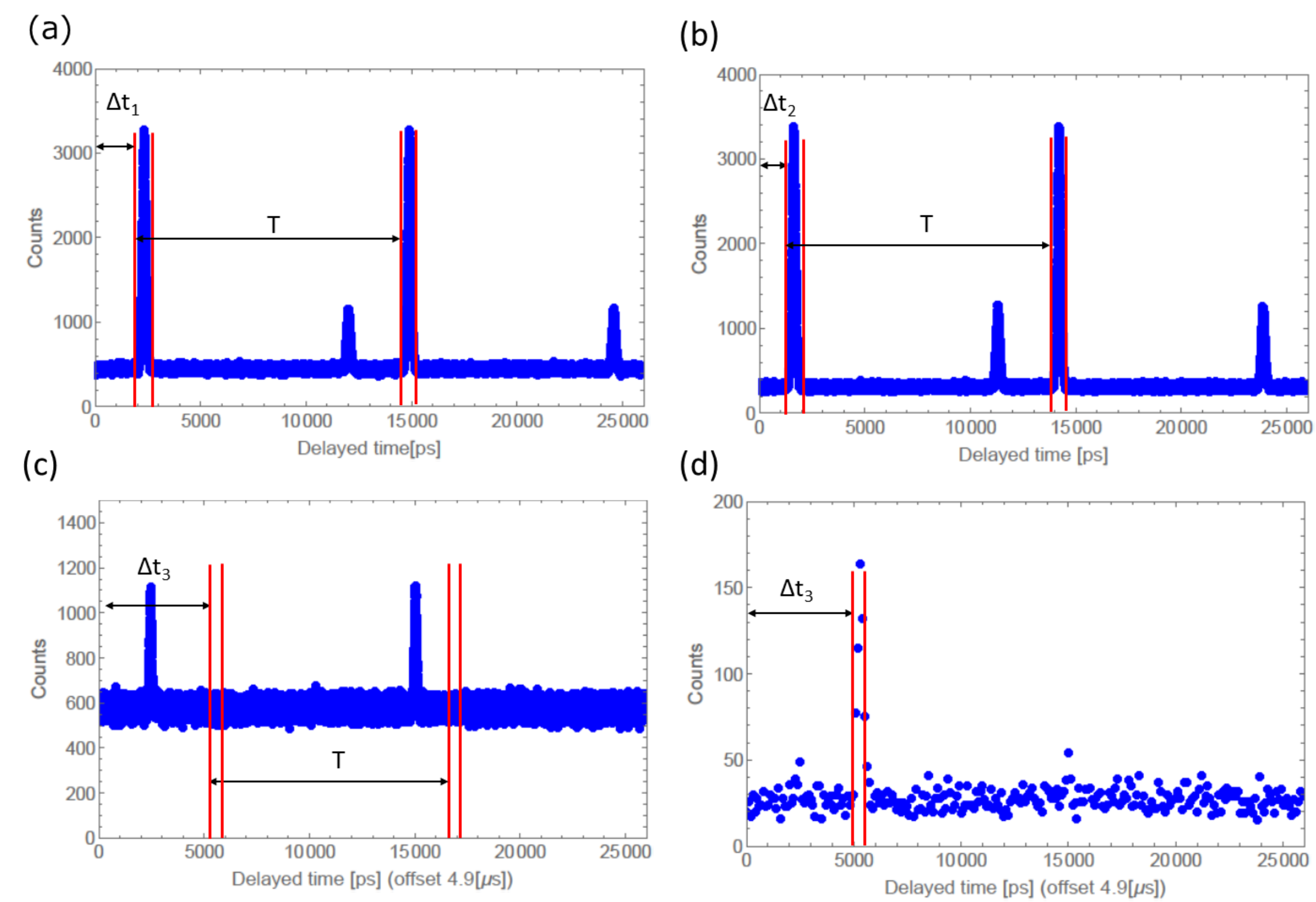}
    \caption{The delayed counts of (a) $\rm{D_{A'}}$, (b)$\rm{D_{R''}}$ and (c) $\rm{D_{B^{\prime}}}$ conditioned by the electric signal from pulse laser.
The figure (d) shows the delayed counts of $\rm{D_{B^{\prime}}}$ conditioned by the photon detection at $\rm{D_{A'}}$ with delayed time $\Delta t_1$.
Each point is integrated for 100ps in (d).}
    \label{Coincidence}
\end{figure}

\subsection{Data processing}
Here, we describe the details of data processing.
The three-fold coincidence events are obtained as follows.
The electric signal from the pulse laser is used as a start signal, and the electric signals from $\rm{D_{A'}}, \rm{D_{R'}}$, and $\rm{D_{B^{\prime}}}$ are used as stop signals.
The histograms of the stop signals are shown in Figs.~\ref{Coincidence} (a), (b) and (c).
In Figs.~\ref{Coincidence} (a) and (b), a higher peak and a lower peak are observed in every 12.5ns (=1/80MHz).
The higher peak, which is the desired one, is obtained when photon R from Bob transmits the HBS$\rm{_A}$.
On the other hand, the lower peak as the unwanted peak is obtained when photon R is reflected by the HBS$\rm{_A}$ and passes through the Sagnac loop.
Such unwanted peaks are also observed in Fig.~\ref{Coincidence} (c) by the photon R coming back to Bob's side after traveling the Sagnac loop.
In the experiment, we removed these unwanted detection events by proper settings of the coincidence time windows.
Since the entangled photon pairs are emitted continuously, when we postselect the higher peaks in Fig. 3~(a) or Fig.3~(b), an additional peak as a counterpart of the photon pair appears at $\Delta t\u{3}$ in the delayed signal at $\rm{D_{B'}}$.
As an example, we show the two-fold coincidence counts between $\rm{D_{A'}}$ and $\rm{D_{B^{\prime}}}$ in Fig.~\ref{Coincidence}~(d).
We extract the successful events of the DFS protocol by summing up the three-fold coincidence events among $\rm{D_{A'}}$, $\rm{D_{R''}}$ and $\rm{D_{B^{\prime}}}$ with timings $\Delta t\u{1}$, $\Delta t\u{2}$ and $\Delta t\u{3}$, respectively.
The widths of the coincidence windows are set to be 300~ps for $\rm{D_{A'}}$ and $\rm{D_{R''}}$, and 100~ps for $\rm{D_{B^{\prime}}}$ respectively.

\section{conclusion}
In conclusion, we demonstrated the DFS protocol with counter-propagating coherent light pulse generated by a fully independent source at telecom band over 1~km of SMFs.
In the demonstration, we employed cw-pulse hybrid system, i.e. a pulsed coherent light source and a continuously emitting photon pair source with time-resolved coincidence measurement, which enables us to perform the DFS scheme without synchronizing independent sources.
The fidelity of the shared photon pair was 0.75$\pm$0.10, which indicates that entanglement is protected by the DFS after traveling in the SMFs.
In practical use, the communication distance of the DFS protocol is limited by the stability of the optical fibre since the noise for the signal photon and the counter-propagating coherent light pulse must satisfy the collective condition.
In this regard, a field experiment shows 
the fluctuations in a 67-km optical fibre are much slower than the round trip time~\cite{Zhou03}.
We believe that our result will be useful for realizing faithful entanglement distribution over a long distance.

\section{acknowledgment}
This work was supported by CREST, JST JPMJCR1671; MEXT/JSPS KAKENHI Grant Number JP18H04291, JP18K13483, and JP18K13487.
K.M. is supported by JSPS KAKENHI No. 19J10976 and Program for Leading Graduate Schools:
Interactive Materials Science Cadet Program.

\section{Supplementary Material}

\begin{figure}[t]
\begin{center}
\scalebox{0.2}{\includegraphics{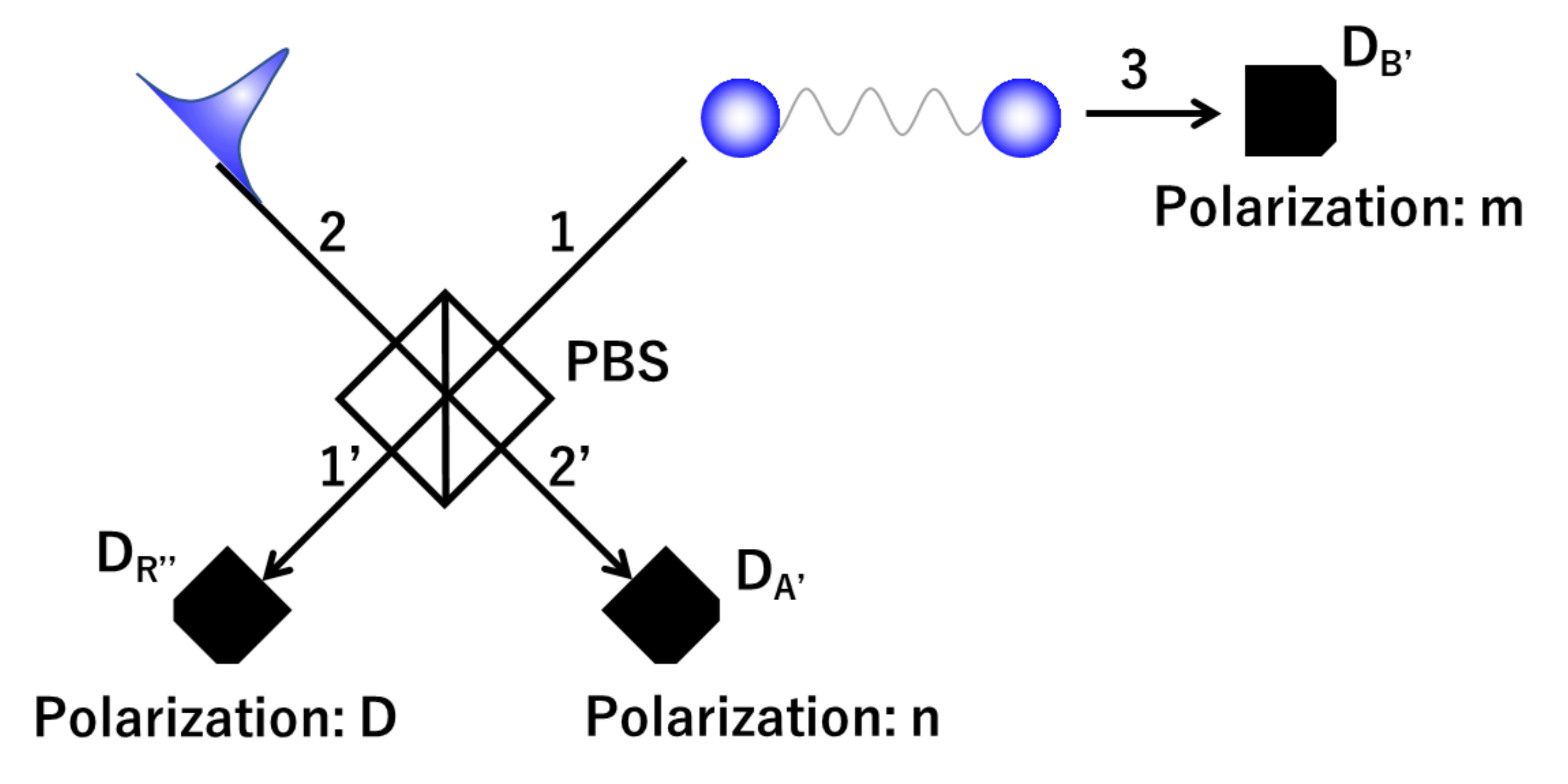}}
 \caption{
   The sketch of our theoretical model. The detection signal of a photon in the mode~3 is used to herald photon in the mode~1. 
   The QPC is performed on the coherent light pulse in the mode~2 and photon in the mode~1. 
   \label{fig:model}}
\end{center}
\end{figure}

In this supplemental material,
we discuss the reason for the degradation of the fidelity of the final state based on the theoretical analysis in Ref.~{\cite{Tsujimoto2018}}. 
We define the coincidence probability among $\mathrm{D\u{R''}}$ with D-polarization, $\mathrm{D\u{B'}}$ with $m$-polarization and $\mathrm{D\u{A'}}$ with $n$-polarization by $P_{mn}$
for $m,n=$H, V, D, A, R, L, where A, R, and L represent antidiagonal polarization, right circular polarization and left circular polarization, respectively. 
Assuming that the single detection probability at D$\u{B'}$
does not depend on the measured polarization $m$, 
we define polarization correlation visibilities of the final state as: 
$V\u{Z}:=|P\u{HH}+P\u{VV}-P\u{HV}-P\u{VH}|/(P\u{HH}+P\u{VV}+P\u{HV}+P\u{VH})$, 
$V\u{X}:=|P\u{DD}+P\u{AA}-P\u{DA}-P\u{AD}|/(P\u{DD}+P\u{AA}+P\u{DA}+P\u{AD})$ 
and $V\u{Y}:=|P\u{RR}+P\u{LL}-P\u{RL}-P\u{LR}-|/(P\u{RR}+P\u{LL}+P\u{RL}+P\u{LR})$. 
Then, the fidelity of the final state is given by 
\begin{equation}
F=(1+V\u{Z}+V\u{X}+V\u{Y})/4. 
\end{equation}
In the following, we estimate the influence of the multiple photon generation at the SPDC photon source and the coherent laser pulse on the polarization correlation visibilities of the final state,
while assuming the perfect indistinguishability. 
The model is shown in Fig~\ref{fig:model}.
The states input to the quantum parity check from modes 1 and 2 are the state heralded by photon detection at D$\u{B'}$
and D-polarized coherent state 
(average photon number: $s_2$, intensity correlation function: 1).
We assume that when an $m$-polarized photon in mode 3 is detected at D$\u{B'}$, it heralds $m^{\ast}$-polarized signal photons
(average photon number: $s_1$, intensity correlation function: $g^{(2)}_{s}$) in mode 1,
while multi-pair emission from the SPDC source produces
$m'$-polarized noise photons
(average photon number: $n_1$, intensity correlation function: 2)
which have no correlation or phase relations with signal photons.
$m^{\ast}$ is complex conjugate to the $m$-polarization and $m'$-polarized state is orthogonal to $m^{\ast}$-polarized state. 
Here, we assume that $s_1$, $n_1$ and $g^{(2)}_{s}$ do not depend on heralding polarization $m$.
As in our experiment, we assume that the overall transmittance of the system including the 
quantum efficiencies of $\U{D}\u{A'}$, $\U{D}\u{R''}$ and $\U{D}\u{B'}$ 
are much less than 1 such that the detection probabilities are proportional to the photon number in the detected mode. 

Under the above assumptions and conditions, we first derive $V_Z$. 
The coincidence probability $P_{m{\rm H}}$ is given by
\begin{eqnarray}
P_{m{\rm H}}&=&\eta\u{1'}\eta\u{2'}
\expect{\hat{b}^{\dagger}\u{1'D}\hat{b}^{\dagger}\u{2'H}\hat{b}\u{1'D}\hat{b}\u{2'H}}\\
&=&\frac{1}{2}\eta\u{1'}\eta\u{2'}
(\expect{\hat{b}^{\dagger}\u{1'H}\hat{b}^{\dagger}\u{2'H}\hat{b}\u{1'H}\hat{b}\u{2'H}}\nonumber \\
&&+\expect{\hat{b}^{\dagger}\u{1'V}\hat{b}^{\dagger}\u{2'H}\hat{b}\u{1'V}\hat{b}\u{2'H}})
\label{equation:P'mn},
\end{eqnarray}
where $\eta_{i}$ is the transmittance of the system in modes $i=1', 2'$.
Here, we introduced the annihilation and creation operators 
$\hat{b}_{i,m}$ and $\hat{b}^{\dagger}_{i,m}$
of the $m$-polarized photons in the output modes of PBS for $i=1',2'$.
They satisfy the commutation relation as $[\hat{b}_{ik},\hat{b}^{\dagger}_{i'k'}]=\delta_{ii'}\delta_{kk'}$. 
Similarly, we define the annihilation and creation operators
$\hat{a}_{i,m}$ and $\hat{a}^{\dagger}_{i,m}$
of the $m$-polarized input modes, 
which satisfy the commutation relation as $[\hat{a}_{jk},\hat{a}^{\dagger}_{j'k'}]=\delta_{jj'}\delta_{kk'}$.
Using a unitary operator $\hat{U}$ of the PBS satisfying 
$\hat{U}\hat{b}^\dagger_{1'\U{H}(1'\U{V})}\hat{U}^\dagger
=\hat{a}^\dagger_{1\U{H}(2\U{V})}$ and 
$\hat{U}\hat{b}^\dagger_{2'\U{H}(2'\U{V})}\hat{U}^\dagger
=\hat{a}^\dagger_{2\U{H}(1\U{V})}$, 
Eq.~(\ref{equation:P'mn}) is transformed into 
\begin{eqnarray}
P_{m{\rm H}}&=&\frac{1}{2}\eta\u{1'}\eta\u{2'}\left
(\expect{\hat{a}^{\dagger}\u{1H}\hat{a}^{\dagger}\u{2H}\hat{a}\u{1H}\hat{a}\u{2H}}+
\expect{\hat{a}^{\dagger}\u{1V}\hat{a}^{\dagger}\u{1H}\hat{a}\u{1V}\hat{a}\u{1H}}\right)
\label{equation:P'mn2}. 
\end{eqnarray}

Rewriting the operators in mode~2 of Eq.~(\ref{equation:P'mn2}) in terms of the polarizations $\U{D}$ and $\U{A}$, we obtain
\begin{eqnarray}
P_{m{\rm H}}
&=&\frac{1}{4}\eta\u{1'}\eta\u{2'}
\Bigl\{\expect{\hat{a}^{\dagger}\u{1H}\hat{a}^{\dagger}\u{2D}\hat{a}\u{1H}\hat{a}\u{2D}}
+\expect{\hat{a}^{\dagger}\u{1H}\hat{a}^{\dagger}\u{2A}\hat{a}\u{1H}\hat{a}\u{2A}} \nonumber \\
&&+\frac{1}{2}(\expect{\hat{a}^{\dagger}\u{2D}\hat{a}^{\dagger}\u{2D}\hat{a}\u{2D}\hat{a}\u{2D}}
+\expect{\hat{a}^{\dagger}\u{2A}\hat{a}^{\dagger}\u{2A}\hat{a}\u{2A}\hat{a}\u{2A}})\Bigr\}.
\label{equation:PHH}
\end{eqnarray}
For calculating $P\u{HH}$, we substitute
$\expect{\hat{a}^{\dagger}\u{1H}\hat{a}\u{1H}}=s\u{1}$,
$\expect{\hat{a}^{\dagger}\u{2D}\hat{a}\u{2D}}=s\u{2}$,
$\expect{\hat{a}^{\dagger}\u{2A}\hat{a}\u{2A}}=0$,  
$\expect{\hat{a}^{\dagger}\u{2D}\hat{a}^{\dagger}\u{2D}\hat{a}\u{2D}\hat{a}\u{2D}}
=s^2_2$ and $\expect{\hat{a}^{\dagger}\u{2A}\hat{a}^{\dagger}\u{2A}\hat{a}\u{2A}\hat{a}\u{2A}}
=0$ leading to
\begin{eqnarray}
P\u{HH}
&=& \frac{1}{4}\eta\u{1'}\eta\u{2'}
\left(s\u{1} s\u{2}
+\frac{1}{2}s\u{2}^2\right).
\label{PHH}
\end{eqnarray}
Similarly, $P\u{VH}$ is calculated by substituting
$\expect{\hat{a}^{\dagger}\u{1H}\hat{a}\u{1H}}=n\u{1}$,
$\expect{\hat{a}^{\dagger}\u{2D}\hat{a}\u{2D}}=s\u{2}$,
$\expect{\hat{a}^{\dagger}\u{2A}\hat{a}\u{2A}}=0$,  
$\expect{\hat{a}^{\dagger}\u{2D}\hat{a}^{\dagger}\u{2D}\hat{a}\u{2D}\hat{a}\u{2D}}
=s^2_2$ and $\expect{\hat{a}^{\dagger}\u{2A}\hat{a}^{\dagger}\u{2A}\hat{a}\u{2A}\hat{a}\u{2A}}
=0$ as
\begin{eqnarray}
P\u{VH} &=& \frac{1}{4}\eta\u{1'}\eta\u{2'}
\left(n\u{1} s\u{2}
+\frac{1}{2}s\u{2}^2\right).
\label{PHV}
\end{eqnarray}

In the same manner, $P_{m{\rm V}}$ is calculated as 
\begin{eqnarray}
P_{m{\rm V}} &=& \eta\u{1'}\eta\u{2'}
\expect{\hat{b}^{\dagger}\u{1'D}\hat{b}^{\dagger}\u{2'V}
\hat{b}\u{1'D}\hat{b}\u{2'V}}\nonumber \\
&=&\frac{1}{2}\eta\u{1'}\eta\u{2'}
\expect{(\hat{b}^{\dagger}\u{1'H}+\hat{b}^{\dagger}\u{1'V})\hat{b}^{\dagger}\u{2'V}
(\hat{b}\u{1'H}+\hat{b}\u{1'V})\hat{b}\u{2'V}}\nonumber \\
&=&\frac{1}{2}\eta\u{1'}\eta\u{2'}
\expect{(\hat{a}^{\dagger}\u{1H}+\hat{a}^{\dagger}\u{2V})\hat{a}^{\dagger}\u{1V}
(\hat{a}\u{1H}+\hat{a}\u{2V})\hat{a}\u{1V}}\nonumber \\
&=&\frac{1}{2}\eta\u{1'}\eta\u{2'}
(\expect{\hat{a}^{\dagger}\u{1H}\hat{a}^{\dagger}\u{1V}\hat{a}\u{1H}\hat{a}\u{1V}}
+\expect{\hat{a}^{\dagger}\u{2V}\hat{a}^{\dagger}\u{1V}\hat{a}\u{2V}\hat{a}\u{1V}})\nonumber \\
&=&\frac{1}{2}\eta\u{1'}\eta\u{2'}
\Bigl\{\expect{\hat{a}^{\dagger}\u{1H}\hat{a}^{\dagger}\u{1V}\hat{a}\u{1H}\hat{a}\u{1V}}
+\frac{1}{2}(\expect{\hat{a}^{\dagger}\u{2D}\hat{a}^{\dagger}\u{1V}\hat{a}\u{2D}\hat{a}\u{1V}}\nonumber \\
&&+\expect{\hat{a}^{\dagger}\u{2A}\hat{a}^{\dagger}\u{1V}\hat{a}\u{2A}\hat{a}\u{1V}})
\Bigr\}.
\label{equation:PVV}
\end{eqnarray}
For calculating $P\u{VV}$, we substitute
$\expect{\hat{a}^{\dagger}\u{1V}\hat{a}\u{1V}}=s_1$,
$\expect{\hat{a}^{\dagger}\u{1H}\hat{a}\u{1H}}=n_1$,
$\expect{\hat{a}^{\dagger}\u{2D}\hat{a}\u{2D}}=s_2$ and 
$\expect{\hat{a}^{\dagger}\u{2A}\hat{a}\u{2A}}=0$
leading to
\begin{eqnarray}
  P\u{VV} &=& \frac{1}{2}\eta\u{1'}\eta\u{2'}
  \left(n\u{1} s\u{1}
  +\frac{1}{2}s\u{1} s\u{2}\right).
  \label{PVV}
\end{eqnarray}
$P\u{HV}$ is calculated by substituting
$\expect{\hat{a}^{\dagger}\u{1V}\hat{a}\u{1V}}=n_1$,
$\expect{\hat{a}^{\dagger}\u{1H}\hat{a}\u{1H}}=s_1$,
$\expect{\hat{a}^{\dagger}\u{2D}\hat{a}\u{2D}}=s_2$ and 
$\expect{\hat{a}^{\dagger}\u{2A}\hat{a}\u{2A}}=0$
as
\begin{eqnarray}
  P\u{HV} &=& \frac{1}{2}\eta\u{1'}\eta\u{2'}
  \left(n\u{1} s\u{1}
  +\frac{1}{2}n\u{1} s\u{2}\right).
  \label{PVH}
\end{eqnarray}

From Eq.~(\ref{PHH}), (\ref{PHV}), (\ref{PVV}) and (\ref{PVH}), the visibility $V\u{z}$ is calculated to be
\begin{eqnarray}
  V\u{z} &=& \frac{2\chi\u{s}(1-\chi\u{n})}
  {4\chi\u{n}+2\chi\u{s}(\chi\u{n}+1)+\chi^2\u{s}},
  \label{Vz}
\end{eqnarray}
where $\chi\u{s}=s\u{2}/s\u{1}$ and $\chi\u{n}=n\u{1}/s\u{1}$.\\

Next, we calculate $V\u{x}$.
The coincidence probability $P_{m{\rm D}}$ is given by
\begin{eqnarray}
P_{m{\rm D}} &=& \eta\u{1'}\eta\u{2'}
\expect{\hat{b}^{\dagger}\u{1'D}\hat{b}^{\dagger}\u{2'D}\hat{b}\u{1'D}\hat{b}\u{2'D}}\nonumber \\
&=&\frac{1}{16}\eta\u{1'}\eta\u{2'}
(\expect{\hat{a}^{\dagger}\u{1D}\hat{a}^{\dagger}\u{1D}\hat{a}\u{1D}\hat{a}\u{1D}}
+\expect{\hat{a}^{\dagger}\u{1A}\hat{a}^{\dagger}\u{1A}\hat{a}\u{1A}\hat{a}\u{1A}}\nonumber \\
&&+\expect{\hat{a}^{\dagger}\u{2D}\hat{a}^{\dagger}\u{2D}\hat{a}\u{2A}\hat{a}\u{2D}}
+\expect{\hat{a}^{\dagger}\u{2A}\hat{a}^{\dagger}\u{2A}\hat{a}\u{2A}\hat{a}\u{2A}}) \nonumber \\
&&+ 4 \expect{\hat{a}^{\dagger}\u{1D}\hat{a}^{\dagger}\u{2D}\hat{a}\u{1D}\hat{a}\u{2D}}
+4 \expect{\hat{a}^{\dagger}\u{1A}\hat{a}^{\dagger}\u{2A}\hat{a}\u{1A}\hat{a}\u{2A}}.
\label{equation:PDD}
\end{eqnarray}
By substituting
$\expect{\hat{a}^{\dagger}\u{1D}\hat{a}\u{1D}}=s\u{1}$,
$\expect{\hat{a}^{\dagger}\u{1A}\hat{a}\u{1A}}=n\u{1}$,
$\expect{\hat{a}^{\dagger}\u{2D}\hat{a}\u{2D}}=s\u{2}$,
$\expect{\hat{a}^{\dagger}\u{2A}\hat{a}\u{2A}}=0$,
$\expect{\hat{a}^{\dagger}\u{1D}\hat{a}^{\dagger}\u{1D}\hat{a}\u{1D}\hat{a}\u{1D}}
=g^{(2)}\u{s}s\u{1}^2$,
$\expect{\hat{a}^{\dagger}\u{1A}\hat{a}^{\dagger}\u{1A}\hat{a}\u{1A}\hat{a}\u{1A}}
=2 n\u{1}^2$,
$\expect{\hat{a}^{\dagger}\u{2D}\hat{a}^{\dagger}\u{2D}\hat{a}\u{2D}\hat{a}\u{2D}}
=s\u{2}^2$,
\begin{eqnarray}
P\u{DD}
&=&\frac{1}{16}\eta\u{1'}\eta\u{2'}
\left\{s\u{2}^2
+ s\u{1}^2 g^{(2)}\u{s}
+ 2 n\u{1}^2 
+ 4  s\u{1}s\u{2}\right\}.
\label{PDD}
\end{eqnarray}
Similarly, $P\u{AD}$ is calculated by substituting
$\expect{\hat{a}^{\dagger}\u{1D}\hat{a}\u{1D}}=n\u{1}$,
$\expect{\hat{a}^{\dagger}\u{1A}\hat{a}\u{1A}}=s\u{1}$,
$\expect{\hat{a}^{\dagger}\u{2D}\hat{a}\u{2D}}=s\u{2}$,
$\expect{\hat{a}^{\dagger}\u{2A}\hat{a}\u{2A}}=0$,
$\expect{\hat{a}^{\dagger}\u{1D}\hat{a}^{\dagger}\u{1D}\hat{a}\u{1D}\hat{a}\u{1D}}
=2 n\u{1}^2$,
$\expect{\hat{a}^{\dagger}\u{1A}\hat{a}^{\dagger}\u{1A}\hat{a}\u{1A}\hat{a}\u{1A}}
=g^{(2)}\u{s}s\u{1}^2$,
and $\expect{\hat{a}^{\dagger}\u{2D}\hat{a}^{\dagger}\u{2D}\hat{a}\u{2D}\hat{a}\u{2D}}
=s\u{2}^2$
as
\begin{eqnarray}
P\u{AD} &=&\frac{1}{16}\eta\u{1'}\eta\u{2'}
\left\{s\u{2}^2
+ s\u{1}^2 g^{(2)}\u{s}
+ 2 n\u{1}^2
+ 4  n\u{1}s\u{2}\right\}.
\label{PAD}
\end{eqnarray}
In the same way, $P_{m{\rm A}}$ is calculated as
\begin{eqnarray}
P_{m{\rm A}} &=& \eta\u{1'}\eta\u{1'}
\expect{\hat{b}^{\dagger}\u{1'D}\hat{b}^{\dagger}\u{2'A}\hat{b}\u{1'D}\hat{b}\u{2'A}}\nonumber \\
&=&\frac{1}{16}\eta\u{1'}\eta\u{2'}
(\expect{\hat{a}^{\dagger}\u{1D}\hat{a}^{\dagger}\u{1D}\hat{a}\u{1D}\hat{a}\u{1D}}
+\expect{\hat{a}^{\dagger}\u{1A}\hat{a}^{\dagger}\u{1A}\hat{a}\u{1A}\hat{a}\u{1A}}\nonumber \\
&&+\expect{\hat{a}^{\dagger}\u{2D}\hat{a}^{\dagger}\u{2D}\hat{a}\u{2D}\hat{a}\u{2D}}
+\expect{\hat{a}^{\dagger}\u{2A}\hat{a}^{\dagger}\u{2A}\hat{a}\u{2A}\hat{a}\u{2A}}\nonumber \\
&&+4\expect{\hat{a}^{\dagger}\u{2D}\hat{a}^{\dagger}\u{1A}\hat{a}\u{2D}\hat{a}\u{1A}}
+4\expect{\hat{a}^{\dagger}\u{1D}\hat{a}^{\dagger}\u{2A}\hat{a}\u{1D}\hat{a}\u{2A}}).
\end{eqnarray}
For calculating $P\u{AA}$, we substitute
$\expect{\hat{a}^{\dagger}\u{1D}\hat{a}\u{1D}}=s\u{1}$,
$\expect{\hat{a}^{\dagger}\u{1A}\hat{a}\u{1A}}=n\u{1}$,
$\expect{\hat{a}^{\dagger}\u{2D}\hat{a}\u{2D}}=s\u{2}$,
$\expect{\hat{a}^{\dagger}\u{2A}\hat{a}\u{2A}}=0$,
$\expect{\hat{a}^{\dagger}\u{1D}\hat{a}^{\dagger}\u{1D}\hat{a}\u{1D}\hat{a}\u{1D}}
=g^{(2)}\u{1}s\u{1}^2$,
$\expect{\hat{a}^{\dagger}\u{1A}\hat{a}^{\dagger}\u{1A}\hat{a}\u{1A}\hat{a}\u{1A}}
=2 n\u{1}^2$,
$\expect{\hat{a}^{\dagger}\u{2D}\hat{a}^{\dagger}\u{2D}\hat{a}\u{2D}\hat{a}\u{2D}}
=s\u{2}^2$,
leading to
\begin{eqnarray}
P\u{AA}
&=&\frac{1}{16}\eta\u{1'}\eta\u{2'}
\left\{s\u{2}^2
+ s\u{1}^2 g^{(2)}\u{s}
+ 2 n\u{1}^2
+ 4 s\u{1}s\u{2}\right\}.
\label{PAA}
\end{eqnarray}
Similarly, $P\u{DA}$ is calculated by substituting
$\expect{\hat{a}^{\dagger}\u{1D}\hat{a}\u{1D}}=n\u{1}$,
$\expect{\hat{a}^{\dagger}\u{1A}\hat{a}\u{1A}}=s\u{1}$,
$\expect{\hat{a}^{\dagger}\u{2D}\hat{a}\u{2D}}=s\u{2}$,
$\expect{\hat{a}^{\dagger}\u{2A}\hat{a}\u{2A}}=0$,
$\expect{\hat{a}^{\dagger}\u{1D}\hat{a}^{\dagger}\u{1D}\hat{a}\u{1D}\hat{a}\u{1D}}
=2 n\u{1}^2$,
$\expect{\hat{a}^{\dagger}\u{1A}\hat{a}^{\dagger}\u{1A}\hat{a}\u{1A}\hat{a}\u{1A}}
=g^{(2)}\u{s}s\u{1}^2$,
$\expect{\hat{a}^{\dagger}\u{2D}\hat{a}^{\dagger}\u{2D}\hat{a}\u{2D}\hat{a}\u{2D}}
=s\u{2}^2$,
as
\begin{eqnarray}
P\u{DA} &=&\frac{1}{16}\eta\u{1'}\eta\u{2'}
\left\{s\u{2}^2
+ s\u{1}^2 g^{(2)}\u{s}
+ 2 n\u{1}^2 
+ 4  n\u{1}s\u{2}\right\}.
\label{PDA}
\end{eqnarray}
From Eq.~(\ref{PDD}), (\ref{PAD}), (\ref{PAA}) and (\ref{PDA}),
the visibility $V\u{x}$ is calculated to be
\begin{eqnarray}
  V\u{x} &=& \frac{2\chi\u{s}(1-\chi\u{n})}
  {\chi\u{s}^2 +2 \chi^2\u{n}  + g^{(2)}\u{s} + 2(\chi\u{n}+1)\chi\u{s}}.
  \label{Vx}
\end{eqnarray}

The coincidence probabilities $P\u{RR},P\u{RL},P\u{LR}$ and $P\u{LL}$
are calculated in a similar manner, which leads to $V\u{x} = V\u{y}$.
As a result, we obtain
\begin{eqnarray}
F &=& \frac{1+V\u{z}+2V\u{x}}{4}.
\label{F}
\end{eqnarray}

In our experiment, the parameters $g^{(2)}_s$, $\chi_s$, and $\chi_n$ are obtained as follows.
The intensity correlation function $g^{(2)}_s=0.098$ is obtained by running another experiment.
The average number of the $m^{\ast}$-polarized heralded photon $s_1'$ is estimated from the experimental data.
The detection probability of H-polarized photon at D$\u{R''}$ conditioned on 
the H-polarized photon-detection at D$\u{B'}$ is given by $s'_1:=C\u{HH}(\U{D}\u{R''}\cap \U{D}\u{B'})/S\u{H}(\U{D}\u{B'})=6.0 \times 10^{-3}$, 
where $C\u{HH}(\U{D}\u{R''} \cap \U{D}\u{B'})$ is the two-fold coincidence count between D$\u{R''}$ with H-polarization and D$\u{B'}$ with H-polarization, 
and $S\u{H}(\U{D}\u{B'})$ is the single detection count at D$\u{B'}$ with H-polarization.
The relation between $s_1$ and $s'_1$ is given by $s_1=s'_1/\eta_d$, 
where $\eta_d$ is the system transmittance after the PBS including the quantum efficiency of the SSPD. 
Similarly, $n_1$ is given by $n_1=n'_1/\eta_d$, where $n'_1:=C\u{HV}(\U{D}\u{R''} \cap \U{D}\u{B'})/S\u{H}(\U{D}\u{B'}) = 1.1 \times 10^{-4}$.
Then, $\chi_n$ is calculated to be $\chi_n=n_1/s_1=n'_1/s'_1=0.018$.
Finally, we estimate the average photon number of the coherent light pulse $s_2$. This is given by $s_2=s'_2/\eta_d$, where $s'_2=2S\u{H}(\U{D}\u{R''})/f = 1.7 \times 10^{-3}$.
Here $f$ is the repetition rate of the mode lock laser. The factor 2 arises from the fact that D-polarized coherent light pulse is divided into two at the PBS.  
Then, $\chi_s$ is given by $\chi_s=s_2/s_1=s'_2/s'_1=0.28$.

Using those parameters, Eq.~(\ref{Vz}), (\ref{Vx}) and (\ref{F}),
we obtain $V_z = 0.76$, $V_x = 0.74$ and $F = 0.81$.


\begin{thebibliography}{10}
    \newcommand{\enquote}[1]{``#1''}
    
    \bibitem{kimble2008}
    H.~J. Kimble, \enquote{The quantum internet,} Nature \textbf{453}, 1023--1030
      (2008).
    
    \bibitem{Wehner18}
    S.~Wehner, D.~Elkouss, and R.~Hanson, \enquote{Quantum internet: A vision for
      the road ahead,} Science \textbf{362} (2018).
    
    \bibitem{Gisin2002}
    N.~Gisin, G.~Ribordy, W.~Tittel, and H.~Zbinden, \enquote{Quantum
      cryptography,} Rev. Mod. Phys. \textbf{74}, 145--195 (2002).
    
    \bibitem{Lo2014}
    H.-K. Lo, M.~Curty, and K.~Tamaki, \enquote{Secure quantum key distribution,}
      Nature Photonics \textbf{8}, 595--604 (2014).
    
    \bibitem{Sangouard2011}
    N.~Sangouard, C.~Simon, H.~de~Riedmatten, and N.~Gisin, \enquote{Quantum
      repeaters based on atomic ensembles and linear optics,} Rev. Mod. Phys.
      \textbf{83}, 33--80 (2011).
    
    \bibitem{Cirac1999}
    J.~I. Cirac, A.~K. Ekert, S.~F. Huelga, and C.~Macchiavello,
      \enquote{Distributed quantum computation over noisy channels,} Phys. Rev. A
      \textbf{59}, 4249--4254 (1999).
    
    \bibitem{Broadbent2009}
    A.~Broadbent, J.~Fitzsimons, and E.~Kashefi, \enquote{Universal blind quantum
      computation,} in \enquote{2009 50th Annual IEEE Symposium on Foundations of
      Computer Science,}  (2009), pp. 517--526.
    
    \bibitem{Lidar2003}
    D.~A. Lidar and K.~Birgitta~Whaley, \emph{Decoherence-Free Subspaces and
      Subsystems} (Springer Berlin Heidelberg, Berlin, Heidelberg, 2003), pp.
      83--120.
    
    \bibitem{Kwiat2000}
    P.~G. Kwiat, A.~J. Berglund, J.~B. Altepeter, and A.~G. White,
      \enquote{Experimental verification of decoherence-free subspaces,} Science
      \textbf{290}, 498--501 (2000).
    
    \bibitem{Walton2003}
    Z.~D. Walton, A.~F. Abouraddy, A.~V. Sergienko, B.~E.~A. Saleh, and M.~C.
      Teich, \enquote{Decoherence-free subspaces in quantum key distribution,}
      Phys. Rev. Lett. \textbf{91}, 087901 (2003).
    
    \bibitem{Boileau2004}
    J.-C. Boileau, D.~Gottesman, R.~Laflamme, D.~Poulin, and R.~W. Spekkens,
      \enquote{Robust polarization-based quantum key distribution over a
      collective-noise channel,} Phys. Rev. Lett. \textbf{92}, 017901 (2004).
    
    \bibitem{Bourennane2004}
    M.~Bourennane, M.~Eibl, S.~Gaertner, C.~Kurtsiefer, A.~Cabello, and
      H.~Weinfurter, \enquote{Decoherence-free quantum information processing with
      four-photon entangled states,} Phys. Rev. Lett. \textbf{92}, 107901 (2004).
    
    \bibitem{Boileau22004}
    J.-C. Boileau, R.~Laflamme, M.~Laforest, and C.~R. Myers, \enquote{Robust
      quantum communication using a polarization-entangled photon pair,} Phys. Rev.
      Lett. \textbf{93}, 220501 (2004).
    
    \bibitem{Yamamoto2005}
    T.~Yamamoto, J.~Shimamura, \ifmmode \mbox{\c{S}}\else~\c{S}\fi{}. K.~\"Ozdemir,
      M.~Koashi, and N.~Imoto, \enquote{Faithful qubit distribution assisted by one
      additional qubit against collective noise,} Phys. Rev. Lett. \textbf{95},
      040503 (2005).
    
    \bibitem{Chen2006}
    T.-Y. Chen, J.~Zhang, J.-C. Boileau, X.-M. Jin, B.~Yang, Q.~Zhang, T.~Yang,
      R.~Laflamme, and J.-W. Pan, \enquote{Experimental quantum communication
      without a shared reference frame,} Phys. Rev. Lett. \textbf{96}, 150504
      (2006).
    
    \bibitem{Prevedel2007}
    R.~Prevedel, M.~S. Tame, A.~Stefanov, M.~Paternostro, M.~S. Kim, and
      A.~Zeilinger, \enquote{Experimental demonstration of decoherence-free one-way
      information transfer,} Phys. Rev. Lett. \textbf{99}, 250503 (2007).
    
    \bibitem{Yamamoto2007}
    T.~Yamamoto, R.~Nagase, J.~Shimamura, \ifmmode \mbox{\c{S}}\else~\c{S}\fi{}.
      K.~\"Ozdemir, M.~Koashi, and N.~Imoto, \enquote{Experimental ancilla-assisted
      qubit transmission against correlated noise using quantum parity checking,}
      New Journal of Physics \textbf{9}, 191 (2007).
    
    \bibitem{Yamamoto2008}
    T.~Yamamoto, K.~Hayashi, \ifmmode \mbox{\c{S}}\else~\c{S}\fi{}. K.~\"Ozdemir,
      M.~Koashi, and N.~Imoto, \enquote{Robust photonic entanglement distribution
      by state-independent encoding onto decoherence-free subspace,} Nature
      Photonics \textbf{2}, 488--491 (2008).
    
    \bibitem{IkutaDFS2011}
    R.~Ikuta, Y.~Ono, T.~Tashima, T.~Yamamoto, M.~Koashi, and N.~Imoto,
      \enquote{Efficient decoherence-free entanglement distribution over lossy
      quantum channels,} Phys. Rev. Lett. \textbf{106}, 110503 (2011).
    
    \bibitem{Kumagai2013}
    H.~Kumagai, T.~Yamamoto, M.~Koashi, and N.~Imoto, \enquote{Robustness of
      quantum communication based on a decoherence-free subspace using a
      counter-propagating weak coherent light pulse,} Phys. Rev. A \textbf{87},
      052325 (2013).
    
    \bibitem{Ikuta2017}
    R.~Ikuta, S.~Nozaki, T.~Yamamoto, M.~Koashi, and N.~Imoto,
      \enquote{Experimental demonstration of robust entanglement distribution over
      reciprocal noisy channels assisted by a counter-propagating classical
      reference light,} Scientific Reports \textbf{7}, 4819 (2017).
    
    \bibitem{Tsujimoto2018}
    Y.~Tsujimoto, M.~Tanaka, N.~Iwasaki, R.~Ikuta, S.~Miki, T.~Yamashita, H.~Terai,
      T.~Yamamoto, M.~Koashi, and N.~Imoto, \enquote{High-fidelity entanglement
      swapping and generation of three-qubit ghz state using asynchronous telecom
      photon pair sources,} Scientific Reports \textbf{8}, 1446 (2018).
    
    \bibitem{Pittman2001}
    T.~B. Pittman, B.~C. Jacobs, and J.~D. Franson, \enquote{Probabilistic quantum
      logic operations using polarizing beam splitters,} Phys. Rev. A \textbf{64},
      062311 (2001).
    
    \bibitem{Tsujimoto2017}
    Y.~Tsujimoto, Y.~Sugiura, M.~Tanaka, R.~Ikuta, S.~Miki, T.~Yamashita, H.~Terai,
      M.~Fujiwara, T.~Yamamoto, M.~Koashi, M.~Sasaki, and N.~Imoto, \enquote{High
      visibility hong-ou-mandel interference via a time-resolved coincidence
      measurement,} Opt. Express \textbf{25}, 12069--12080 (2017).
    
    \bibitem{Miki2013}
    S.~Miki, T.~Yamashita, H.~Terai, and Z.~Wang, \enquote{High performance
      fiber-coupled nbtin superconducting nanowire single photon detectors with
      gifford-mcmahon cryocooler,} Opt. Express \textbf{21}, 10208--10214 (2013).
    
    \bibitem{James2001}
    D.~F.~V. James, P.~G. Kwiat, W.~J. Munro, and A.~G. White, \enquote{Measurement
      of qubits,} Phys. Rev. A \textbf{64}, 052312 (2001).
    
    \bibitem{Rehacek2007}
    J.~\ifmmode \check{R}\else \v{R}\fi{}eh\'a\ifmmode~\check{c}\else \v{c}\fi{}ek,
      Z.~Hradil, E.~Knill, and A.~I. Lvovsky, \enquote{Diluted maximum-likelihood
      algorithm for quantum tomography,} Phys. Rev. A \textbf{75}, 042108 (2007).
    
    \bibitem{Miki2019}
    S.~{Miki}, S.~{Miyajima}, M.~{Yabuno}, T.~{Yamashita}, T.~{Yamamoto},
      N.~{Imoto}, R.~{Ikuta}, R.~A. {Kirkwood}, R.~H. {Hadfield}, and H.~{Terai},
      \enquote{Timing jitter characterization of the sfq coincidence circuit by
      optically time-controlled signals from sspds,} IEEE Transactions on Applied
      Superconductivity \textbf{29}, 1--4 (2019).
    
    \bibitem{Aktas16}
    D.~Aktas, B.~Fedrici, F.~Kaiser, T.~Lunghi, L.~Labonté, and S.~Tanzilli,
      \enquote{Entanglement distribution over 150 km in wavelength division
      multiplexed channels for quantum cryptography,} Laser \& Photonics Reviews
      \textbf{10}, 451--457 (2016).
    
    \bibitem{Zhou03}
    C.~Zhou, G.~Wu, X.~Chen, and H.~Zeng, \enquote{“plug and play” quantum key
      distribution system with differential phase shift,} Applied Physics Letters
      \textbf{83}, 1692--1694 (2003).
    
    \end{thebibliography}
\end{document}